\PassOptionsToPackage{table,xcdraw}{xcolor}
\documentclass[acmsmall]{acmart}

\AtBeginDocument{%
  \providecommand\BibTeX{{%
    \normalfont B\kern-0.5em{\scshape i\kern-0.25em b}\kern-0.8em\TeX}}}

\usepackage{todonotes}
\usepackage{xspace}
\usepackage{url}
\usepackage{listings}
\usepackage{tikz}
\usepackage{subcaption}
\usepackage{multirow}
\usepackage{tablefootnote}
\usepackage[most]{tcolorbox}
\usepackage[capitalise]{cleveref}
\usepackage{makecell} 
\usepackage{adjustbox}
\usepackage{enumitem}

\definecolor{samsungblue}{HTML}{034c9d}
\newcommand{\name}{\textsc{AutoFL}\xspace}
\newcommand*\circled[1]{\tikz[baseline=(char.base)]{
    \node[shape=circle,draw,inner sep=0.3pt] (char) {#1};}}

\definecolor{verygood}{HTML}{77DD76}
\definecolor{verybad}{HTML}{FF6962}
\definecolor{good}{HTML}{BDE7BD}
\definecolor{bad}{HTML}{FFB6B3}

\begin{document}

\title[A Quantitative and Qualitative Evaluation of LLM-Based Explainable Fault Localization]{A Quantitative and Qualitative Evaluation of LLM-Based Explainable Fault Localization}

\author{Sungmin Kang}
\authornote{Both authors contributed equally to this research.}
\email{sungmin.kang@kaist.ac.kr}
\affiliation{%
  \institution{KAIST}
  \city{Daejeon}
  \country{South Korea}
}

\author{Gabin An}
\authornotemark[1]
\email{gabin.an@kaist.ac.kr}
\affiliation{%
  \institution{KAIST}
  \city{Daejeon}
  \country{South Korea}
}

\author{Shin Yoo}
\email{shin.yoo@kaist.ac.kr}
\affiliation{%
  \institution{KAIST}
  \city{Daejeon}
  \country{South Korea}
}

\renewcommand{\shortauthors}{Kang, An, and Yoo}

\begin{abstract}
  Fault Localization (FL), in which a developer seeks to identify
  which part of the code is malfunctioning and needs to be fixed, is a recurring challenge in debugging. To reduce developer
  burden, many automated FL techniques have been proposed. However, prior work has
  noted that existing techniques fail to provide rationales for the suggested locations, hindering developer adoption of these techniques.
  With this in mind, we propose \name, a Large Language Model (LLM)-based FL technique that generates an explanation of
  the bug along with a suggested fault location. \name prompts an LLM to use function calls to
  navigate a repository, so that it can effectively localize faults over a large software repository
  and overcome the limit of the LLM context length. Extensive experiments on 798 real-world bugs in Java and Python reveal \name~improves method-level acc@1 by up to 233.3\% over
  baselines.
  Furthermore, developers were interviewed on their impression of \name-generated explanations, showing that developers generally liked the natural language explanations of \name, and that they preferred reading a few, high-quality explanations instead of many.
\end{abstract}



\maketitle

\section{Introduction}
\label{sec:intro}
Fault Localization (FL) is the task of identifying which part of a software system is responsible for a bug. FL has been surveyed to take the majority of human debugging time: for example, B{\"o}hme et al. report that bug diagnosis (which includes localization) took 66\% of debugging time in their human study~\cite{Bohme2017DbgBench}.
As a result, FL has been widely researched, with techniques from slicing-based FL~\cite{Agrawal1995SlicingFL}, to others that use causality analysis~\cite{Kucuk2021UniVal} or model the propagation of erroneous states~\cite{Zeng2022SmartFL}. 

Despite the promise of automated FL techniques, they have not been widely adopted by practitioners. We are unaware of the industrial adoption of FL systems, unlike the case for automated program repair (APR)~\cite{SapFix2018Marginean, Winter2022DevAPR}. 
Looking at the various expectations that developers have of FL tools outlined by Kochhar et al.~\cite{Kochhar2016FLExpect}, one aspect that developers emphasize is \emph{providing rationales}: almost 90\% of developers agreed that the capability of an FL technique to generate rationales is important.
In contrast, to the best of our knowledge, there are few FL techniques that can explain why a particular location is the likely culprit in an accessible manner. Consequently, an \emph{explainable} FL technique, one that can explain both the bug and the reason why the suggested location is linked to the buggy behavior, would take FL techniques one step closer to practitioner adoption.

In this paper, we present \name, an automated FL technique that uses Large Language Models (LLMs) to not only suggest potential bug locations, but also to provide explanations on how the bug occurred and why the suggested location is suspicious. LLMs are difficult to apply to FL in a realistic manner, as a critical input for FL is the entire software repository, which can span thousands to millions of lines of code, whereas the context length of LLMs is generally limited - for example, the largest LLMs at the time of writing can process 32,000 \emph{tokens} at once, much less lines. To solve this problem, we allow LLMs to navigate the source code by allowing them to call functions that return information on covered classes, their covered functions, and the implementation and documentation for any covered function. Given a prompt asking for the LLM to find the answer using the available functions, the LLM invokes a series of function calls to gather relevant information, and reaches a conclusion on how the bug manifested, which we treat as the \emph{explanation} of the bug. 
By construction, \name has many benefits when compared to existing FL techniques: it only requires a single failing test, it can deal with multiple programming languages seamlessly, and critically can generate explanations for developers.%

We perform thorough experiments to evaluate both the FL capability of \name and the quality of its explanations. Our quantitative evaluation on the FL performance of \name using real-world bug datasets from Java and Python indicates that when using the GPT-3.5 language model, it could achieve comparable or superior performance to existing techniques. For example, when comparing against the Ochiai SBFL technique, \name improves on its method-level acc@1 score by 19.7\% on the Defects4J benchmark and 166.7\% on the BugsInPy benchmark; using GPT-4, \name could perform even better, outperforming Ochiai by 233.3\% on BugsInPy. By repeating the execution of \name~\cite{wang2023selfconsistency}, we also demonstrate that we can estimate the level of \emph{confidence} \name has in its results, providing an avenue for reducing false positives when presenting results to developers.

Meanwhile, evaluating the quality of explanations generated by \name is a difficult, yet important aspect when assessing \name. First, manual analysis of 300 sampled explanations from 60 bugs reveals that about 20\% of generated explanations accurately describe how the bug happened, while 56.7\% of all bugs had an accurate description. 
Critically, we presented the explanations of \name to professional developers, and performed interviews on what they liked and disliked about the explanations.
Our key findings are that developers were generally supportive of explaining FL results and bugs, with many suggesting that the natural language descriptions of the failure were helpful, but that redundant content and incorrect fix suggestions could lead them astray. 
Developers additionally indicated the desire to see a few explanations instead of many.%

These results suggest that %
it would be helpful to automatically identify which explanations are helpful. We present the results of our follow-up preliminary experiments which estimate the quality of explanations based on the quality of tests and patches made with an LLM conditioned on those explanations. We find a positive correlation with the correctness of explanations, providing hints as to how explanations could be automatically assessed for developer usage. 

To summarize, our contributions are:
\begin{itemize}[leftmargin=15pt]
    \item We introduce \name, an LLM-based FL technique that overcomes the input limitations of LLMs by allowing them to autonomously retrieve relevant portions of the code, and generates an explanation of how the bug happened before suggesting a fault location.
    \item We rigorously evaluate the FL performance of \name, and find that it can outperform FL baselines over two large real-world bug benchmarks, and that it can indicate when its answer is likely to be correct, potentially reducing developer hassle with false positives.
    \item We evaluate the explanations generated by \name, and find that accurate explanations could be generated for more than half of all bugs, while developers had positive views towards the explanations generated by \name.
\end{itemize}

The remainder of the paper is organized as follows. \cref{sec:background} provides the academic background, and \cref{sec:approach} describes how \name works. The evaluation setup is described in \cref{sec:expr_setup}, with the results in \cref{sec:results}. \cref{sec:future_dirs} describes our preliminary experiments to automatically identify helpful explanations, \cref{sec:ttv} lays out threats to validity, and \cref{sec:conclusion} concludes.

\section{Background}
\label{sec:background}
This section provides the background and research context. 

\subsection{LLM Tool Use}
By integrating chain-of-thought prompting~\cite{Wei2022ChainOT} with the output of tools, ReAct~\cite{yao2022react} demonstrated that LLMs are capable of interacting with tools to achieve better performance on tasks. Since then, LLM interaction with external tools has been widely explored. For example, HuggingGPT~\cite{shen2023hugginggpt} has LLMs compose computer vision pipelines by dynamically integrating the results of various computer vision models together. 
LLM tool use has also been explored in software engineering, notably for program repair: Xia et al.~\cite{xia2023conversational} integrated test feedback into the prompt, while Kang et al.~\cite{kang2023explainable} allows LLMs to invoke a debugger.

Recent iterations of OpenAI's LLMs have embraced this change and added a feature named function calling~\cite{OpenAIFunctionCalling}. This capability enables users to provide function descriptions to the LLM, which can respond with JSON data requesting a function call, complete with arguments required for calling the function, on the digression of the LLM. 
For instance, to answer a user inquiry about the current weather, an LLM may call a function that retrieves the weather of a particular location, which would be processed in an automated manner and presented to the LLM so that it can provide a coherent response.\footnote{While this is notably implemented by the OpenAI API, tools such as LangChain~\cite{Chase_LangChain_2022} allow other LLMs to incorporate functions as well; preliminary results using Llama2~\cite{touvron2023llama2} with \name are given in our supplementary material~\cite{autoflsupplementary}.}
In this context, we aim to define a set of functions that the LLM can employ to gather necessary information for debugging.

\begin{table}[t]
\caption{A comparison of existing FL techniques with \name. The precision of SBFL, MBFL, and IRFL was recalculated based on the artifacts of Zou et al.~\cite{Zou2019CombineFL}; for other techniques, precision comes from the corresponding papers. Wu et al.~\cite{wu2023LLMFL} only evaluate statement-level FL, so their precision could not be compared.}
\scalebox{0.8}{
\begin{tabular}{rccccc}
\Xhline{1pt}
              & Required Artifact                                              & Prec@5 on D4J & Time                                                & Multilang.                & Rationale                   \\
\Xhline{0.5pt}
SBFL          & \cellcolor{good}Test suite                             & \cellcolor{bad}61\%~\cite{Zou2019CombineFL}          & \cellcolor{good}minutes                     & \cellcolor{good}yes & \cellcolor{bad}no  \\
MBFL          & \cellcolor{good}Test suite                             & \cellcolor{bad}54\%~\cite{Zou2019CombineFL}          & \cellcolor{bad}hours                       & \cellcolor{good}yes & \cellcolor{bad}no  \\
IRFL          & \cellcolor{good}Bug report                             & \cellcolor{verybad}3\%\tablefootnote{While Zou et al. report this performance in their paper, their replication package did not include IRFL results.}~\cite{Zou2019CombineFL}          & \cellcolor{verygood}seconds                     & \cellcolor{good}yes & \cellcolor{bad}no  \\
CombineFL~\cite{Zou2019CombineFL} & \cellcolor{bad}All of the above                       & \cellcolor{bad}69\%~\cite{Zou2019CombineFL}          & \cellcolor{bad}hours                       & \cellcolor{bad}no  & \cellcolor{bad}no  \\
DeepRL4FL~\cite{Li2021rl4fl} & \cellcolor{good}Test suite                       & \cellcolor{good}79\%~\cite{Li2021rl4fl}          & \cellcolor{bad}hours                       & \cellcolor{bad}no  & \cellcolor{bad}no  \\
UniVal~\cite{Kucuk2021UniVal}        & \cellcolor{good}Pass/Fail test                         & \cellcolor{good}75\%~\cite{Kucuk2021UniVal}          & \cellcolor{good}minutes\tablefootnote{This was achieved by only instrumenting the buggy class, which would not be possible in practice.}                     & \cellcolor{bad}no  & \cellcolor{bad}no  \\
SmartFL~\cite{Zeng2022SmartFL}       & \cellcolor{good}Pass/Fail test                         & \cellcolor{good}70\%~\cite{Zeng2022SmartFL}          & \cellcolor{good}minutes                     & \cellcolor{bad}no  & \cellcolor{bad}no  \\
Wu et al.~\cite{wu2023LLMFL}     & \multicolumn{1}{l}{\cellcolor{verybad}Buggy method/class} & -             & - & \cellcolor{good}yes & \cellcolor{good}yes \\
AutoFL        & \cellcolor{verygood}Single test                            & \cellcolor{good}Up to 71\%          & \cellcolor{good}minutes                     & \cellcolor{good}yes & \cellcolor{good}yes \\
\Xhline{1pt}
\end{tabular}}
\label{tab:fl_characteristics}
\end{table}

\subsection{Fault Localization}
Fault Localization (FL) is a critical process in debugging that involves identifying specific locations in a program's source code where bugs are present. Automated FL techniques help developers save time, particularly in large codebases, by accurately pinpointing the code locations most likely to be responsible for the target bug. 
We provide a comparison of existing FL work with \name in \cref{tab:fl_characteristics}. Commonly used FL technique families include Spectrum-based FL (SBFL), Information Retrieval-based FL (IRFL), and Mutation-based FL (MBFL)~\cite{Wong2016FLSurvey}. While SBFL techniques are known to be the most effective standalone techniques~\cite{Zou2019CombineFL}, they require coverage data from both passing and failing tests. Meeting this requirement can be challenging for large enterprise software, where coverage measurement can have high computational costs~\cite{Ivankovic2019codecoveragegoogle, Bach2022saphana, kim2003efficient}. Additionally, most FL techniques lack a rationale or explanation in their output, limiting their reliability and practicality in practical debugging. As Kochhar et al.~\cite{Kochhar2016FLExpect} note, rationales for FL are crucial for bug fixing, as clear rationales enable developers to understand why a particular location is identified as the culprit for the bug, helping them incorporate their domain knowledge and make informed decisions. Meanwhile, the recent work of Wu et al.~\cite{wu2023LLMFL} presents a buggy method/class and asks an LLM which location is likely to be buggy and why. 
However, it is difficult to use as a standalone FL technique in practice, as it requires prerequisite knowledge of which method/class is buggy for operation. 

\section{Approach}
\label{sec:approach}

\begin{figure}[t]
    \centering
    \includegraphics[width=0.75\linewidth]{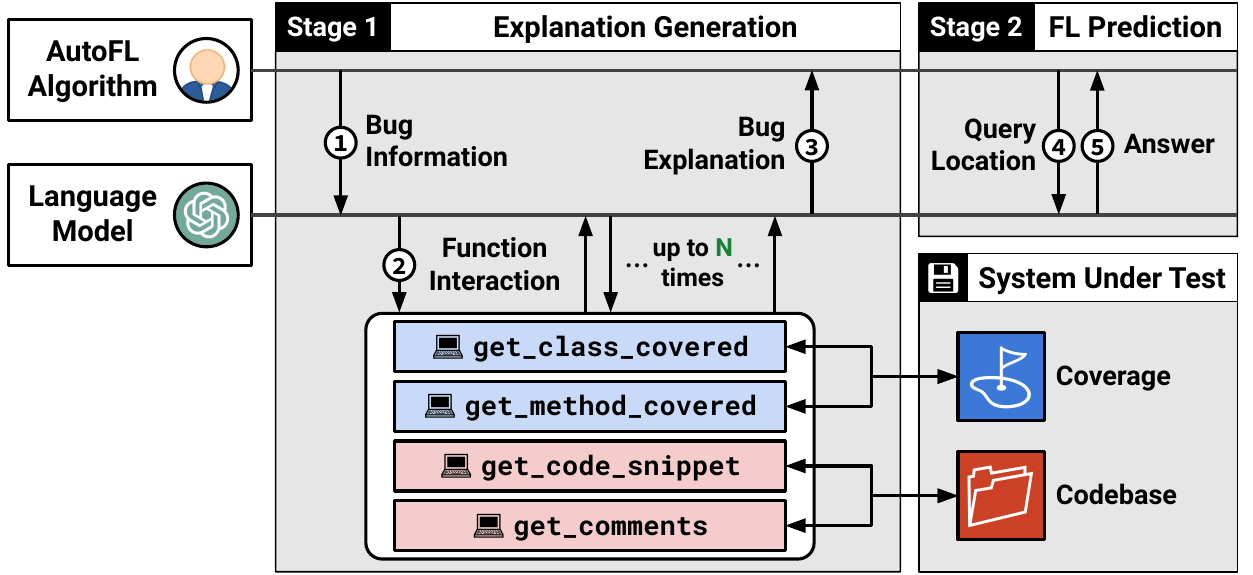}
    \caption{Diagram of \name. Arrows represent an interaction between components; circled numbers indicate the order. Function interactions are made at most N times, where N is a predetermined parameter of \name.}
    \label{fig:diagram}
\end{figure}

In this paper, we introduce \name, a novel \textbf{auto}mated and \textbf{auto}nomous \textbf{FL} technique that harnesses LLMs to localize bugs in software given a single 
failing test. A key advantage of using LLMs is that they can generate natural language \emph{explanations} on why a particular location seems likely to be buggy, as LLMs are adept in both software engineering and natural language processing tasks~\cite{brown2020language, xia2022practical}. Conceptually, we define an `explanation' as any text that helps the developer comprehend the bug or how \name came to a conclusion; to actualize this ideal and make such explanations, we designed \name to provide the logical flow from the root cause to the failure, as was identified a key component for failure explanations by Du et al.~\cite{Du2023}. \name represents a new type of effort in FL research, in the sense that we can now readily generate explanations along with localization results, marking a qualitative difference from prior techniques. %

\name specifically focuses on method-level FL for two main reasons. First, the method granularity is the favored granularity level among developers according to prior work~\cite{Kochhar2016FLExpect}; furthermore, recent work on LLM-based automated debugging techniques often require method-level localization results as a minimum~\cite{xia2022practical, jiang2023impact, wu2023LLMFL}, suggesting that localization on the method level is a recurring problem in LLM-based automated debugging. As mentioned earlier, dealing with large code repositories is a challenge for LLMs, but we tackle this issue by equipping LLMs with custom-designed functions to enable code exploration and relevant information extraction.

An overview of \name is depicted in \cref{fig:diagram}. We employ a two-stage prompting process, where Stage 1 involves inquiring about the root cause of the given failure, and Stage 2 requests output about the fault location.
In Stage 1, 
\circled{1} \name provides a prompt to the LLM containing failing test information and descriptions of available functions for debugging.
\circled{2} The LLM interacts with the provided functions autonomously, to extract the information needed for the debugging of the given failure. 
\circled{3} Based on the gathered information, the LLM generates an explanation about the root cause of the observed failure. 
In Stage 2, 
\circled{4} the LLM is queried to return the culprit location for the bug, and 
\circled{5} the LLM responds by providing the culprit method.
In doing so, we can explicitly acquire both the \textbf{root cause explanation} and the \textbf{bug location}.

\begin{figure}[h]
\begin{lstlisting}[caption={System Prompt for LLM},label={lst:system-prompt},numbers=none]
You are a debugging assistant. You will be presented with a failing test, and tools (functions) to access the source code of the system under test (SUT). Your task is to provide a step-by-step explanation of how the bug occurred, based on the failing test and the information you retrieved using tests about the SUT. You will be given N chances to interact with functions to gather relevant information. An example answer would look like follows.

|<HANDCRAFTED ROOT CAUSE ANALYSIS EXAMPLE>|
\end{lstlisting}
\end{figure}

During the process, the system message\footnote{A system message refers to a message that is used to guide the behavior of the LLM by providing instructions.} 
shown in \cref{lst:system-prompt} is used to guide the LLM in its role as a debugging assistant.
As described before, we aimed for \name to generate the logical flow from fault to failure; the prompt was designed with this in mind, as it asks for a \emph{step-by-step} explanation of how the bug occurred. This has the added benefit that it may also improve the performance of \name, as generating explanations before answering is known to help improve performance~\cite{kojima2023large}.
Furthermore, the handcrafted example is a brief one that describes the cause of a bug in two sentences, nudging
\name towards generating concise explanations.

\begin{figure}[h!]
\begin{lstlisting}[caption={Example prompt from Defects4J Lang-48},label={lst:stage1-prompt}, xleftmargin=0.7cm]
The test `~...EqualsBuilderTest::testBigDecimal()~` failed. The test looks like:

```java
~381 :     public void testBigDecimal() {
382 :         BigDecimal o1 = new BigDecimal("2.0");
383 :         BigDecimal o2 = new BigDecimal("2.00");
385 :         assertTrue(new EqualsBuilder().append(o1, o2).isEquals()); // error occurred here
386 :     }~
```

It failed with the following error message and call stack:
```
~junit.framework.AssertionFailedError
  at ...EqualsBuilderTest::testBigDecimal(EqualsBuilderTest.java:385)~
```
Start by calling the `get_failing_tests_covered_classes` function.
\end{lstlisting}
\end{figure}

\subsection{Stage 1: Generating Root Cause Explanation}

At the outset, \name initiates the FL process by presenting an initial prompt about the failed test to the LLM (\cref{fig:diagram}, \circled{1}). This prompt is automatically generated and includes bug-related information such as the failing test and its error stack trace. \cref{lst:stage1-prompt} shows the prompt template, with the relevant failure details highlighted in \textcolor{samsungblue}{blue}.

First, the prompt specifies the name of the failing test and provides the test code snippet (Lines 1-9). The test code snippet is enclosed within triple backticks to indicate it is a code block. We employ two heuristics to minimize irrelevant content that may confuse the LLM. First, we minimize the snippet to exclusively include statements placed prior to where the failure occurs, which is explicitly marked with the comment \texttt{"// error occurred here"}; as test statements after the failure are not executed, they bear less relevance to the bug than the actually executed test code. In case the failure is within a nested block, the entire outermost statement containing the error is included. Additionally, any preceding assertion statements are detected and removed, as we can be sure that these assertions passed and are thus less likely to be relevant to the bug; indeed, their inclusion could confuse the language model. For example, Line 384 in \cref{lst:stage1-prompt} is an assertion statement that passed which is irrelevant to the actual bug. Hence, it was automatically removed and not visible in \cref{lst:stage1-prompt}. While \name uses a single test in its prompt, note that the prompt can be easily extended to handle multiple failing tests by concatenating the information from each test.

Following the snippet, the prompt presents the failure symptoms consisting of an error message and the stack trace (Lines 11-15). Stack traces are automatically minimized by retaining only the information related to the target repository (e.g., lines related to external libraries are omitted). Additionally, repeated subsequences occurring more than five times are condensed to improve readability and conciseness, which helps with long stack traces, e.g., stack overflow errors.

Finally, the prompt ends with a suggestion that the LLM call the \texttt{get\_covered\_classes} function (Line 16), to encourage the LLM to make use of functions when generating its answer. Preliminary experiments showed the addition of this instruction improved the function call success rate by the LLM. Note that we append this initial function call request and its response to the message chain of \name without requiring an actual call from the LLM.

Along with the prompt, \name provides the four specialized functions for debugging and allows the LLM to \textit{decide} whether to (i) request a function call (\cref{fig:diagram}, \circled{2}), with a limit of at most $N$ function interactions, or (ii) to generate a user-facing bug explanation (\cref{fig:diagram}, \circled{3}), which concludes Stage 1. If the LLM requests a function call, \name executes the function and provides the return value back to the model by adding the result to the message history.
The first two functions, indicated in blue in the diagram, allow the LLM to obtain class-level and method-level coverage information related to the failing test. With these functions, the LLM can narrow down the methods associated with the observed failure, enabling a targeted analysis of the root cause. The latter two functions, denoted with red in the diagram, are general code navigation tools, which take the signature of a method of interest as input and return the code snippet and the relevant documentation (if it exists). While simple, these functions permit significant flexibility for the LLM to explore the repository and access the current implementation and specification of methods. Furthermore, as Shuster et al.~\cite{shuster2021retrieval} note, models tend to be more truthful when augmented with document-retrieving tools; thus, by including functions that retrieve information as a part of \name, we increase the likelihood that \name will generate an accurate description of how the bug happened, as was our initial goal. This is also indirectly reflected in our results in \Cref{sec:results}, which show that without this information retrieval, FL performance drops.
If the LLM provides invalid arguments to these functions, a guidance message aimed at improving subsequent function call requests is returned. For instance, when the LLM submits incomplete method signatures that match multiple existing methods, the following guidance message is returned: \textit{"There are multiple matches to that query. Do you mean any of the following: <candidates>?"} 
Additional details on the implementation of such guidance messages can be found in our artifact.  %

\begin{figure}[h]
\begin{lstlisting}[numbers=none,caption=Prompt to Request the Fault Location,label={lst:stage2-prompt}]
Based on the available information, provide the signatures of the most likely culprit methods for the bug. Your answer will be processed automatically, so make sure to only answer with the accurate signatures of the most likely culprit (in `ClassName.MethodName(ArgType1, ArgType2, ...)` format), without commentary (one per line).
\end{lstlisting}
\end{figure}

\subsection{Stage 2: Pinpointing The Fault Location}

After Stage 1 concludes, the LLM is then prompted to predict the culprit methods based on the available information (\cref{fig:diagram}, \circled{4}), based on the prompt in \cref{lst:stage2-prompt}. In this stage, unlike Stage 1, we enforce the model to respond immediately without making function calls, assuming that the LLM has already (implicitly) identified the fault location(s) in Stage 2.
Finally, the entire FL process concludes with the LLM providing a response pinpointing the potential fault locations (\cref{fig:diagram}, \circled{5}). 

For instance, given the initial prompt about Lang-48 (\cref{lst:stage1-prompt}) in Stage 1, the LLM of \name sequentially made four function calls: first, to obtain the class and methods covered by the failing test, then to retrieve code snippets for two methods of \texttt{EqualsBuilder}, namely \texttt{isEquals()} and \texttt{append(Object, Object)}. The LLM then identifies the root cause of the bug: an erroneous utilization of the equals method for \texttt{BigDecimal} objects,
resulting in a comparison based on references rather than values in the \texttt{append} method. In Stage 2, the LLM specifically suggests \texttt{EqualsBuilder.append(Object, Object)} as buggy, which matches the developer patch location.

\subsection{Finalizing Fault Localization Results}

\label{sec:aggregation}
\begin{figure}[t]
    \centering
    \includegraphics[width=0.62\linewidth]{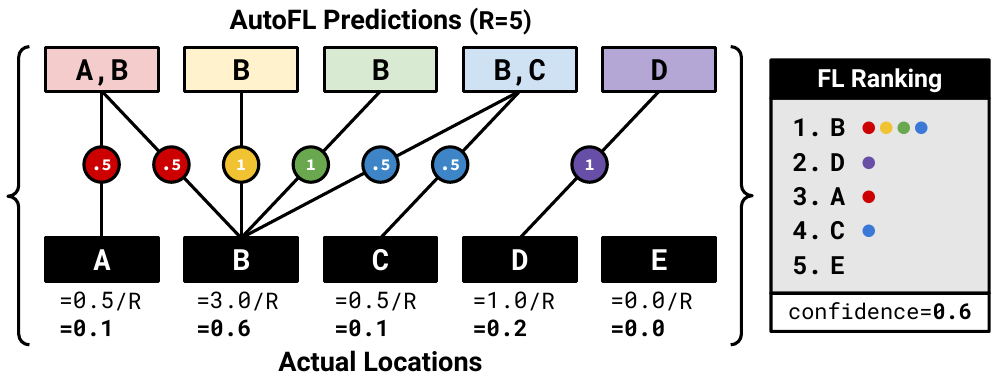}
    \caption{Scoring and ranking candidate methods (depicted as black rectangles) based on five \name prediction outcomes (depicted as colored rectangles). For every prediction outcome, the scores (represented as colored circles) are evenly distributed among all the methods included in that particular outcome.}
    \label{fig:scoring}
\end{figure}

To start, we assign scores to the methods 
by combining the results of $R$ repeated runs of \name.
Specifically, if a final prediction contains a total of $n$ methods, we give a score of $1/n$ to each of these identified methods.\footnote{No scores are distributed in cases where the prediction results are erroneous, or if the entire \name process is interrupted due to other errors.} These individual scores are then averaged over all $R$ predictions. Formally, the score of a method $m$ is defined as:
\begin{equation}
\text{score}(m) = \frac{1}{R}\sum_{k = 1}^{R}(\frac{1}{|r_k|} \cdot [m \in r_k])
\label{eq:score}
\end{equation}
where $r_i$ is the set of predicted methods from the $i$-th run, and $[.]$ is the Iverson bracket which returns 1 when the predicate inside is true, and 0 otherwise.
For example, for method $B$ in \cref{fig:scoring}, there are four \name runs that predict the method as a faulty location. Following \cref{eq:score}, the score for method \texttt{B} would be $(0.5 + 1.0 + 1.0 + 0.5 + 0.0)/5 = 0.6$. After calculating the scores for all methods predicted by \name, we rank them in descending order of scores. For instance, in the given example, the four predicted methods are sorted into \texttt{[B, D, A, C]}, as their scores are \texttt{[0.6, 0.2, 0.1, 0.1]}, respectively. In case of a score tie, we prioritize methods that appeared in earlier predictions over others as a final, arbitrary tiebreaker.

Note that if there are methods with a score of $0$ (i.e., not part of the final \name results) but are covered by the failing tests, e.g., the method \texttt{E} in the given example, we append them to the end of the ranked list to ensure the list includes all methods relevant to the failure. If there are multiple such methods, they are primarily sorted in descending order of the number of failing tests covering each method. In case of ties among them, we give priority to methods that are more frequently mentioned during the function interaction process of \name (\cref{fig:diagram}, \circled{2}), based on the intuition that methods that are inspected by the LLM or related to inspected methods are more likely to be faulty than methods that were never observed in the debugging process.

Finally, after producing a complete ranked list of suspicious methods, e.g., \texttt{[B, D, A, C, E]} in \cref{fig:scoring}, 
our next step involves estimating the confidence in the final predictions. We gauge the level of confidence based on the consistency of the LLM prediction results across multiple iterations, motivated by the previous work on LLM self-consistency~\cite{wang2023selfconsistency}. The intuition is that if the LLM consistently produces similar predictions, we would have more confidence in the results. Therefore, we select the highest score among the methods covered by failing tests, and then use it to determine the confidence score $M$, as follows:
\begin{equation}
\text{confidence} = \max_{m \in M}score(m)
\label{eq:conf}
\end{equation}

Having a confidence measure helps the usability of \name in two main aspects, backed by our results. First, higher confidence tends to correlate with better results; thus, a practitioner may pick a confidence threshold to automatically filter suggestions. Second, users also indicate a preference for techniques that can indicate confidence for efficient use.

\section{Experimental Settings}
\label{sec:expr_setup}

This section details the experimental setup used to evaluate \name. 

\subsection{Research Questions}
\label{sec:rq}

\paragraph{RQ1.} \textbf{How accurately does \name localize faults?}%

\begin{itemize}[leftmargin=15pt]
\item Evaluation Metric: FL performance is evaluated with the acc@k metric which measures the number of bugs for which an actual buggy location was within the top $k$ suggestions of a tool, as previous work suggests developers are only willing to look at a few suggested locations when debugging~\cite{Parnin2011AAD, Kochhar2016FLExpect}, and as this metric is often used by prior work~\cite{Zou2019CombineFL, Xia2019DeepFL, Zeng2022SmartFL}. To deal with ties in the ranking, instead of using the average tiebreaker, we use the ordinal tiebreaker, as we believe it is closer to what a developer would experience when using an FL tool.
\item Baselines: We evaluate the FL capability of \name using a total of 798 reproducible bugs from Defects4J (Java) and BugsInPy (Python); see the dataset details in \Cref{sec:dataset}). For Defects4J, we compare against the best-performing standalone techniques from Zou et al.~\cite{Zou2019CombineFL}, namely the Ochiai~\cite{Abreu2007OchiaiSBFL} and DStar~\cite{Wong2014DStar} SBFL techniques and the Metallaxis~\cite{Papadakis2015Metallaxis} MBFL technique. To ensure consistency with our experimental setup, we recalculated the method-level acc@k metrics for the same set of bugs using the publicly available replication package provided by Zou et al. We also compare against the state-of-the-art standalone FL technique, SmartFL~\cite{Zeng2022SmartFL}. As SmartFL was only evaluated on a subset of Defects4J due to the complex Java features used in Closure~\cite{Zeng2022SmartFL}, the comparison involves only these 222 Defects4J bugs excluding 131 bugs from Closure. Finally, we also compare against the Test-LLM baseline, which predicts the fault location using an LLM based on the failing test and error message alone (i.e., it is \name without function interaction). Meanwhile, for the BugsInPy dataset, we compared against the SBFL results reported by Widyasari et al.~\cite{widyasari2022real}, again recalculating method-level acc@k values for the same set of bugs using their replication package.
\end{itemize}

\paragraph{RQ2.} \textbf{How well does the confidence from \name align with the actual FL performance?}

Before a user decides whether to inspect the generated FL results, it would be beneficial if \name could accurately predict the performance of FL. Therefore, we investigate how well the prediction confidence from \name (defined in \cref{eq:conf}) aligns with FL performance. We calculate Spearman's rank correlation coefficients between the estimated confidence values and the following three widely used ranking metrics (higher is better). These metrics are used as they are defined for individual FL outcomes, whereas acc@k in RQ1 aggregates performance over the entire dataset.

\begin{itemize}[leftmargin=15pt]
\item Precision@1 ($P@1$): The precision at the top rank, meaning it equals 1 if the highest-ranked method is found to be faulty; otherwise, it is 0.
\item Reciprocal Rank ($RR$): The reciprocal rank of the highest-ranked faulty method.
\item Average Precision ($AP$): The average precision values for each rank of the faulty methods.
\end{itemize}

\paragraph{RQ3.} \textbf{How good are the quality of the explanations of \name?}

As an explainable FL technique, it is important to evaluate the quality of the explanations of \name. We anticipate that developers will receive \name-generated FL rankings, with each method on the list paired with a set of explanations identifying why the method is considered faulty. To this end, we manually evaluated 300 explanations from 60 bugs, which were randomly selected from the Defects4J dataset. Explanations were evaluated on the following four criteria, modeled after the bug explanation evaluation criteria from Mahbub et al.~\cite{Mahbub2023}, which are also showcased through real examples in \Cref{fig:category_examples}.
\begin{itemize}[leftmargin=15pt]
    \item Accurate: the explanation contains a detailed description of why the bug occurs, which goes beyond simply explaining the error message.
    \item Imprecise: the explanation contains an inaccurate statement. %
    \item Concise: the explanation succinctly describes why the bug occurs, without extraneous content. %
    \item Useful: the explanation correctly describes how to fix the bug. %
\end{itemize}

Additionally, for the evaluation of the explanations, each explanation was rated as a whole, rather than in fragments. To minimize human error during the evaluation,
two authors manually evaluated each explanation and resolved differences through discussion; these final explanation evaluation results are used in our analysis. We further analyze the relationship between FL confidence and explanation quality - assuming FL results are selectively shown via confidence, \name would be more useful if explanations tend to be of higher quality when FL results are shown.

\paragraph{RQ4.} \textbf{How do professional developers feel about the explanations of \name?}

To evaluate the practical impact of explainable FL techniques, we invited 16 professional developers to use the explanations of \name to debug bugs from the BugsInPy benchmark. In particular, developers are asked to debug two bugs that we selected to be of moderate difficulty following prior work~\cite{Xia2016HarmfulHarmful} for about one hour. Based on this, we present how professional developers feel about the explanations of \name, what still needs to be improved, and in this process identify aspects that future explainable automated debugging techniques should place particular focus on.

\subsection{Experimental Details}
The experimental details for our study are provided.

\subsubsection{Evaluation Dataset}

\begin{table}
\caption{Details of the bug datasets. The terms \#FTs and \#PMs refer to the average number of failing tests and patched methods, respectively. The calculation of patched methods excludes omission bugs. Projects marked with $\dagger$ originate from Defects4J~\cite{Just2014Defects4J}, whereas the remaining ones are sourced from BugsInPy~\cite{Widyasari2020BIP}.}
\scalebox{0.8}{
\begin{tabular}{lrrr|lrrr|lrrr}
\toprule
Project & \#Bugs & \#FTs & \#PMs & Project & \#Bugs & \#FTs & \#PMs & Project & \#Bugs & \#FTs & \#PMs\\
\midrule
PySnooper  & 2 & 1.00 & 2.00 & luigi  & 28 & 1.29 & 1.65 & tornado & 14 & 1.14 & 1.15 \\
ansible  & 11 & 1.27 & 1.45 & matplotlib & 27 & 1.04 & 1.58 & youtube-dl & 42 & 1.00 & 1.39 \\
black  & 22 & 1.05 & 2.43 & pandas & 165 & 1.25 & 1.53 & Chart$\dagger$  & 26 & 3.54 & 1.52 \\
cookiecutter  & 4 & 1.25 & 2.00 & sanic & 3 & 1.00 & 1.00 & Closure$\dagger$  & 131 & 2.63 & 1.25 \\
fastapi  & 16 & 1.94 & 2.42 & scrapy & 38 & 1.45 & 1.17 & Lang$\dagger$  & 64 & 1.92 & 1.38 \\
httpie  & 5 & 1.00 & 1.25 & spacy & 2 & 1.00 & 1.00 & Math$\dagger$ & 106 & 1.66 & 1.32 \\
keras  & 36 & 1.17 & 2.37 & thefuck & 30 & 1.27 & 1.24 & Time$\dagger$ & 26 & 2.85 & 1.56 \\
\bottomrule
\end{tabular}
}
\label{tab:dataset}
\end{table}

\label{sec:dataset}
Our experiment is conducted using a total of 798 bugs from 21 open-source projects as listed in \Cref{tab:dataset}.
We use two different bug benchmarks: Defects4J v1.0~\cite{Just2014Defects4J} (Java), chosen to allow comparison with traditional techniques, and BugsInPy~\cite{Widyasari2020BIP} (Python), to demonstrate that \name can quickly be adapted to other languages as well. For Defects4J, all active 353 bugs in Defects4J v1.0 are used. For BugsInPy, while the core implementation of \name did not need to be updated, we modified the callable function set, as the tests in BugsInPy tended to cover a large number of classes, while Python often includes its comments within the function body. We use the improved BugsInPy dataset by Aguilar et al.~\cite{aguilar2023BetterBIP}; nonetheless, 56 out of 501 bugs are excluded due to reproducibility issues, leaving 445 bugs for consideration. Further details are provided in the supplementary material~\cite{autoflsupplementary}.

\subsubsection{\name Default Configurations}
For each bug, we run \name five times, i.e., $R=5$, using the gpt-3.5-turbo-0613 language model from OpenAI; this number is chosen as we found diminishing returns for $R>5$ in a preliminary study we performed on a subset of the bugs from Defects4J. %
We also set the maximum number of function interactions to $N=10$; this number was chosen as (i) only a small proportion of runs actually use up to 10 runs (less than 10\%), and (ii) preliminary experiments with $N=20$ caused the LLMs to exceed their context length limits often, dropping the performance of \name by half.
While we treat the gpt-3.5-turbo-0613 results as our `main' results and use them in subsequent analysis, we also present the result of running \name two times with the gpt-4-0613 model to show performance when using an improved LLM, and to show that GPT-4 also benefits from our result combination mechanism (\cref{sec:aggregation}).
While \name can function effectively with just one failing test as input, a variety of strategies can be employed for \name when dealing with multiple failing tests. In our experiment, when multiple failing tests are present, we employ distinct failing test cases for each run of \name. More specifically, %
when there are multiple failing test cases (194 bugs out of 798 in total have multiple failing tests), a round-robin approach is adopted, selecting one failing test case for each run.%

\subsubsection{Developer Feedback Details}
Developers of three companies interested in LLM-based FL were invited to use the explanations and FL from \name to debug real bugs from open-source software, and provide qualitative feedback. Specifically, employees of each company who had prior experience with working with the authors internally recruited other employees who were available to participate in the experiment. The average participant had more than five years of experience with software development, and more than 1.8 years of development experience with Python. We made this choice as there are no clear baselines for explainable FL, and as we believed that providing explanation design guidelines from developer experience would be beneficial to future work. Prior to conducting our main study, a pilot study was conducted
to obtain feedback and thus improve the fidelity of our results, as is recommended~\cite{Ko2015SEHumanStudy}, which helped us streamline the process. Overall, we could recruit 16 professional developers who could participate in our study in full. 

In the study, developers were presented with real-world bugs with failing tests, and asked to generate a patch that would fix the bug, similarly to B{\"o}hme et al.~\cite{Bohme2017DbgBench}. We used bugs from the BugsInPy dataset, as Python developers were easier to recruit than C or Java developers. Developers were provided the error message, a failing test, FL results from \name, and critically 10 bug explanations from \name (5 from both GPT-3.5 and GPT-4, and translated from the original English for participant convenience) for their task. The test execution environment was set up so that participants could freely execute tests and perform print debugging. Each participant would first go through a tutorial in which they were asked to fix a simple bug and presented with the ground-truth patch, so that they could get used to the experiment. In turn, developers would debug two bugs over the course of an hour, based solely on the error message and the FL/explanations that \name generated. 
When the developer has made patches for every bug, or when they deem the bug too difficult to debug, we perform a semi-structured interview in which we ask the developer whether FL and FL explanations would be useful in their workflow, what their view of the strengths and weaknesses of \name-generated explanations are, and what they believe an ideal explanation would be. The common themes from developer responses are analyzed and discussed between two authors, and a final multi-label tagging of each developer's response is generated. Based on this tagging, we present the popular themes as results, and present them with quotes from developers. Further details about our developer study can be found in our supplementary material~\cite{autoflsupplementary}.

\section{Results}
\label{sec:results}
This section presents the results of our experiments.

\subsection{RQ1: FL Efficacy}

\begin{figure}[t]
    \centering
    \begin{subfigure}[t]{0.49\linewidth}
        \centering
        \includegraphics[width=\linewidth]{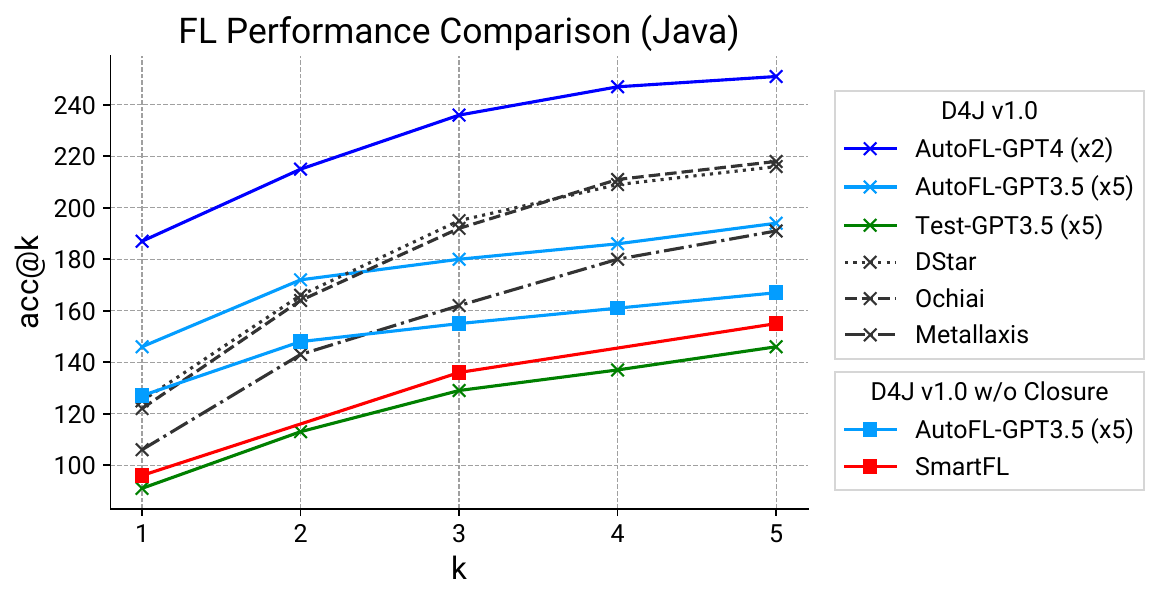}
        \caption{FL evaluation on Defects4J}
        \label{fig:sota_comp_java}
    \end{subfigure}
    \begin{subfigure}[t]{0.49
    \linewidth}
        \centering
        \includegraphics[width=\linewidth]{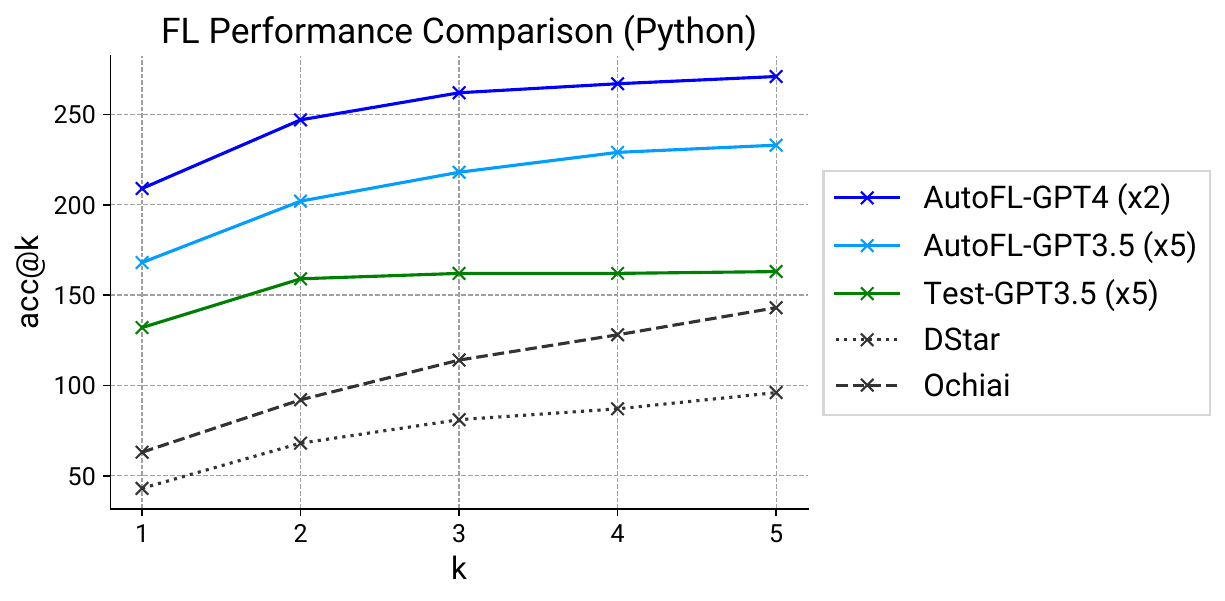}
        \caption{FL evaluation on BugsInPy}
        \label{fig:sota_comp_python}
    \end{subfigure}
    \caption{Performance of various FL techniques on the Defects4J and BugsInPy benchmarks.}
    \label{fig:sota_comparison}
\end{figure}

In \cref{fig:sota_comparison}, \name is compared with other standalone techniques that report method-level performance. For the Defects4J dataset (\cref{fig:sota_comp_java}), the graph shows that \name, when using GPT-3.5 as the language model, outperforms SBFL and MBFL, which were the best-performing standalone FL techniques from Zou et al.~\cite{Zou2019CombineFL}, on the acc@1 measure. Furthermore, \name outperforms the state-of-the-art standalone FL technique SmartFL, which was only evaluated on a subset of Defects4J due to the complex Java features used in Closure~\cite{Zeng2022SmartFL};\footnote{On the same bug dataset, \name still outperformed SmartFL, as shown in the square marker graphs of \cref{fig:sota_comp_java}.} in contrast, \name could be easily expanded to a completely different language, Python. However, the performance of \name at acc@3 and acc@5 lags behind SBFL. This is likely because GPT-3.5 is still a limited LLM; our manual inspection of GPT-3.5 debugging traces reveals that it has difficulty `digging deep' into a repository to find bugs.
This is further confirmed by our experiment with GPT-4: GPT-4 demonstrates stronger performance on reasoning benchmarks than GPT-3.5~\cite{openai2023gpt4}, and similarly \name-GPT4 overcomes the limitations of GPT-3.5 to consistently achieve better performance relative to all baselines up to acc@5.
Comparing against the Test-GPT3.5 baseline, which prompts GPT-3.5 to predict the fault location without any function interaction, \name consistently outperforms it, demonstrating that the function interactions improve the performance of \name.
On the Python benchmark BugsInPy, over which \name was evaluated to demonstrate its multilingual capability, \name-GPT3.5 outperforms SBFL techniques by a substantial margin, as shown in \cref{fig:sota_comp_python}: \name-GPT3.5 and \name-GPT4 improved method-level acc@1 by 166.7\% and 233.3\% when compared with Ochiai. For BugsInPy, \name is performing on a consistent level with Defects4J, while SBFL is significantly worse in BugsInPy, as reported by Widyasari et al.~\cite{widyasari2022real}. Overall, \name can perform consistently even while requiring fewer artifacts from the developer and it can be easily adapted to different languages, indicating the significant potential of LLM-based FL techniques.

\begin{figure}[t]
    \centering
    \begin{subfigure}[t]{0.45\linewidth}
        \centering
        \includegraphics[width=\linewidth]{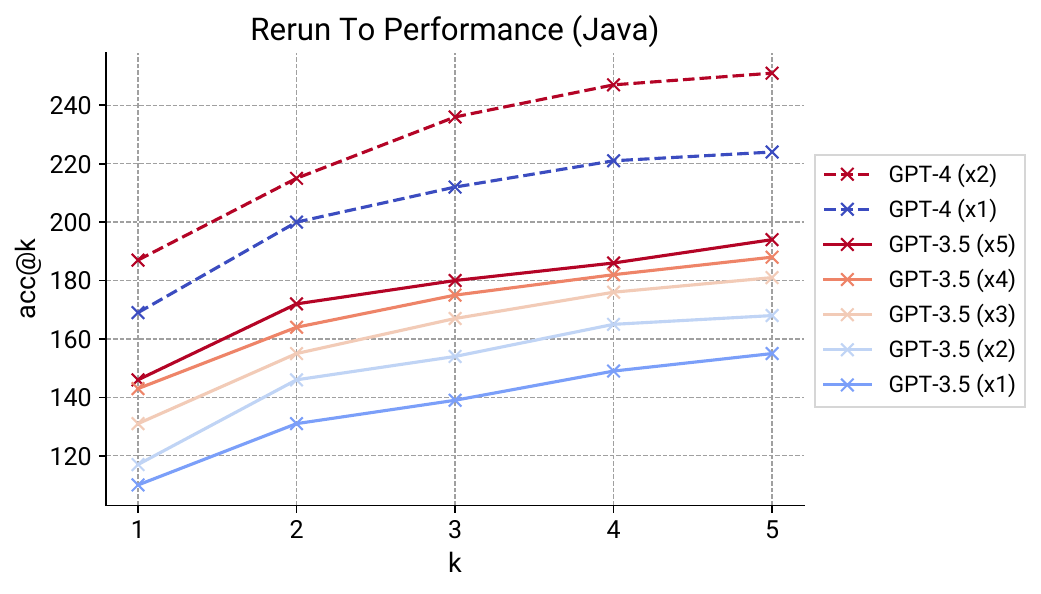}
        \caption{Defects4J}
        \label{fig:merge_java}
    \end{subfigure}
    \begin{subfigure}[t]{0.45\linewidth}
        \centering
        \includegraphics[width=\linewidth]{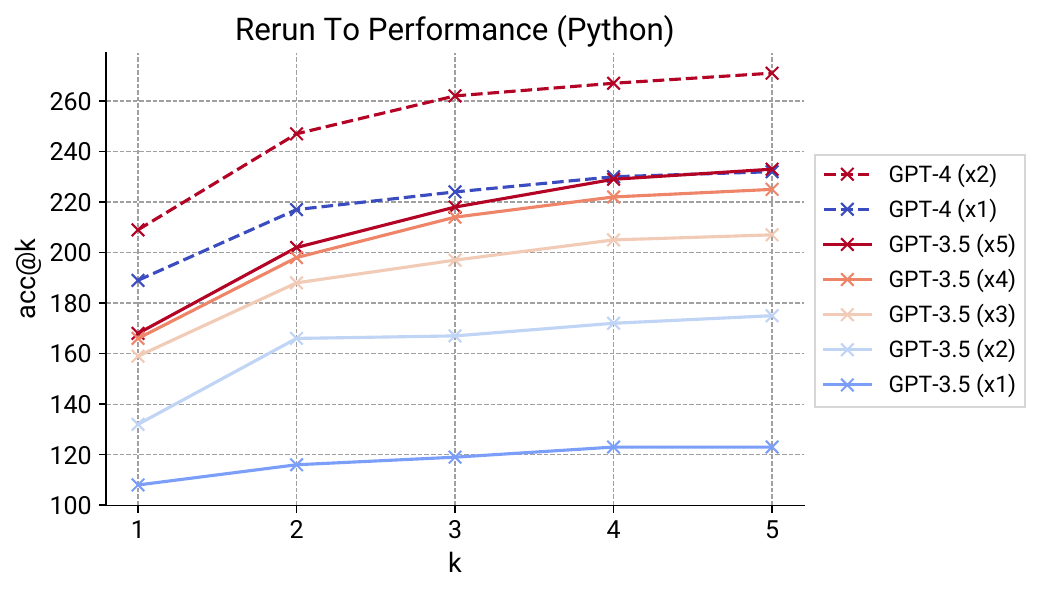}
        \caption{BugsInPy}
        \label{fig:merge_python}
    \end{subfigure}
    \caption{Performance of \name as $R$ increases, for Defects4J and BugsInPy.}
    \label{fig:merge_plot}
\end{figure}

Meanwhile, merging more \name runs to get an aggregate result helps improve performance by a large margin in both Defects4J and BugsInPy, as shown in \cref{fig:merge_plot}. In addition, we observe when using GPT-4, a single run can outperform five aggregated runs of \name-GPT3.5, demonstrating the potential of improved language models contributing to better FL performance.

\begin{table}[t]
\caption{Execution time (seconds) per bug for \name. Prep. denotes the data preparation phase, which involves gathering coverage of the failing test cases and obtaining the snippets of covered code.}
\scalebox{0.8}{
\begin{tabular}{l|l|r|rrr|r|r}
\toprule
Benchmark & LLM &  Prep. (\circled{1}) & $R=1$ & $R=2$ (\circled{2}) & $R=5$ (\circled{2}) & Merge (\circled{3}) & Total (\circled{1}+\circled{2}+\circled{3})\\\midrule
\multirow{2}{*}{Defects4J} & GPT-3.5 &  \multirow{2}{*}{4.87} &  16.40 & - & 82.00 & 0.37 & 87.24 \\
                           & GPT-4   & & 152.48 & 304.96 & - & 0.52 & 310.35 \\\midrule
\multirow{2}{*}{BugsInPy}  & GPT-3.5 & \multirow{2}{*}{23.50} & 34.71 & - & 173.55 & 0.28 & 197.33 \\
                           & GPT-4   & & 65.80 & 131.6 & - & 0.20 & 155.38 \\
\bottomrule
\end{tabular}}
\label{tab:exectime}
\end{table}

\begin{figure}[t]
    \centering
    \begin{subfigure}[t]{0.45\linewidth}
        \centering
        \includegraphics[width=\linewidth]{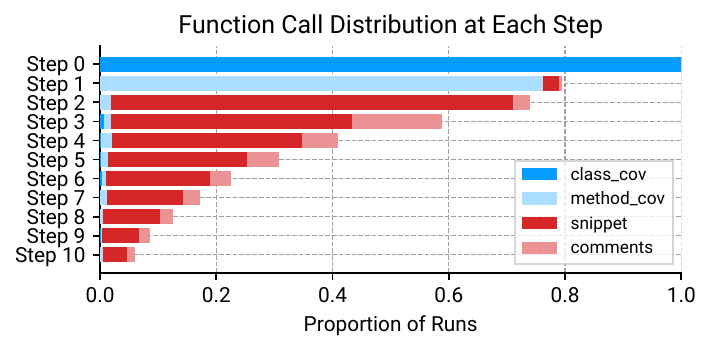}
        \caption{Defects4J}
    \end{subfigure}
    \begin{subfigure}[t]{0.45\linewidth}
        \centering
        \includegraphics[width=\linewidth]{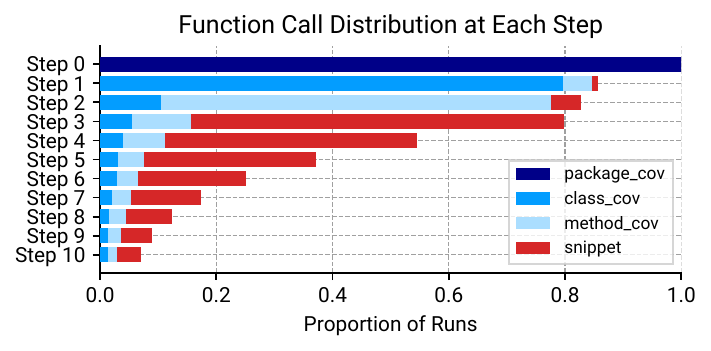}
        \caption{BugsInPy}
    \end{subfigure}
    \caption{Function call distribution for \name-GPT3.5.}
    \label{fig:func_call_dist}
\end{figure}

\begin{figure}[t]
    \centering
    \begin{subfigure}[t]{0.45\linewidth}
        \centering
        \includegraphics[width=\linewidth]{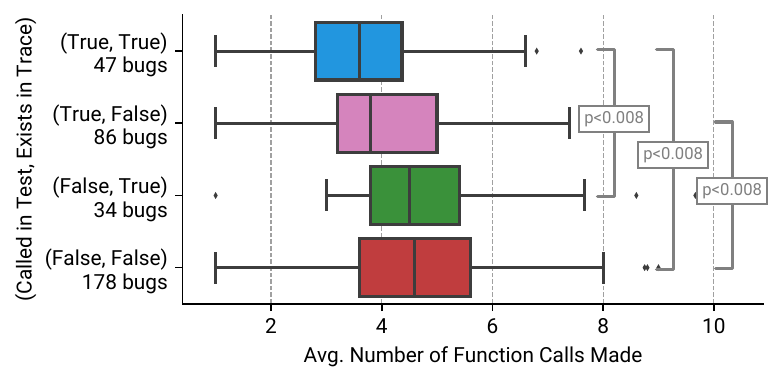}
        \caption{Defects4J}
    \end{subfigure}
    \begin{subfigure}[t]{0.45\linewidth}
        \centering
        \includegraphics[width=\linewidth]{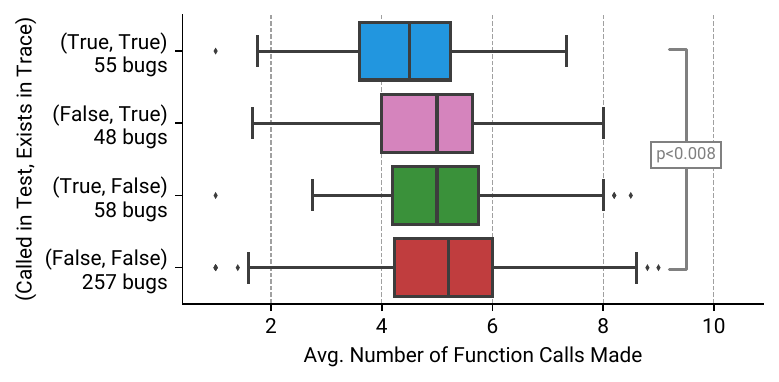}
        \caption{BugsInPy}
    \end{subfigure}
    \caption{Average number of function calls made by \name-GPT3.5, categorized by whether at least one of the faulty methods is called in the failing test case and/or exists in the stack trace.}
    \label{fig:func_call_len}
\end{figure}

Finally, we provide additional information about the characteristics of \name runs. A single run ($R=1$) of \name on a single bug on Defects4J using GPT-3.5 took 16.4 seconds on average, as shown in \cref{tab:exectime}. Together with preparation (4.87s) and result aggregation (0.37s) for a single bug on average, the average runtime of \name with five runs was 87.24 seconds. This is faster than what Zou et al.~\cite{Zou2019CombineFL} report as the time cost of SBFL (112 seconds), indicating that \name can operate as a lightweight FL technique. 
The runtime of \name is dependent on the number of function calls it makes; in our experiments, an average of 5.37 function calls are made to determine the fault location. The type of function calls made with GPT-3.5\footnote{The results of the same analysis for GPT-4 are presented in the supplementary material~\cite{autoflsupplementary}} at each step is presented in \cref{fig:func_call_dist}, and indicates that the length of inference chains is diverse. 

Our analysis further explores how the \textit{localization difficulty} of bugs is linked to the length of function call sequences. The difficulty is heuristically measured by whether at least one of the actual faulty methods is directly mentioned in the failing test code or appears in the stack trace.
If \name were capable of using function calls effectively, \name would perform a deeper search through the function calls for `harder' bugs that do not directly expose the faulty method. As expected, `harder' bugs tended to result in longer chains of function calls (\Cref{fig:func_call_len}). In Defects4J and BugsInPy, \name utilizes on average 0.6 and 0.9 additional function calls respectively to address the hardest bugs, i.e., (False, False), compared to the easiest bugs, i.e., (True, True). A one-way ANOVA confirmed significant differences ($p < 0.05$) in the mean length of function calls across different bug difficulties. Subsequent pairwise t-tests with Bonferroni adjustments applied to the p-value threshold (adjusting it to 0.008 for six combinations) identify which differences are statistically significant, as indicated as gray lines in the figure. %

\begin{tcolorbox}[boxrule=0pt,frame hidden,sharp corners,enhanced,borderline north={1pt}{0pt}{black},borderline south={1pt}{0pt}{black},boxsep=2pt,left=2pt,right=2pt,top=2.5pt,bottom=2pt]
    \textbf{Answer to RQ1:} \name shows comparable or superior performance relative to prior standalone FL techniques with less information, can easily be applied to multiple languages, operates on a timescale of minutes, and adaptively makes function calls depending on bug characteristics.
\end{tcolorbox}

\subsection{RQ2: Predicting FL Accuracy via \name Confidence}

\begin{table}[t]
\caption{Spearman's rank correlation coefficients between \name confidence and FL performance metrics in each benchmark (with `*' denoting p < $0.0001$). \name is rerun 5 times using GPT-3.5.}
\scalebox{0.8}{
\begin{tabular}{lrrr}
\toprule
Correlation with & \textbf{Precision@1} & \textbf{Reciprocal Rank} & \textbf{Average Precision}\\\midrule
Defects4J & +0.57* & +0.67* & +0.70*\\
BugsInPy & +0.52* & +0.50* & +0.49*\\
\bottomrule
\end{tabular}}
\label{tab:conf_corr}
\end{table}

\cref{tab:conf_corr} presents Spearman's rank correlation coefficients that illustrate the relationship between the confidence values and the three FL performance metrics, namely Precision@1, Reciprocal Rank, and Average Precision, described in \cref{sec:rq}. In both benchmark datasets, Defects4J and BugsInPy, we observe statistically significant positive correlations between the \name confidence values and the FL performance metrics. A depiction of performance distribution by confidence bins is presented in the supplementary material~\cite{autoflsupplementary}.
Furthermore, although the correlation is more pronounced within the Defects4J dataset in comparison to the BugsInPy dataset, the correlation with Precision@1 remains relatively consistent across both datasets, with respective values of 0.57 and 0.52. We conjecture that the correlation between Confidence and Precision@1 is more consistent because, unlike other metrics, both are determined solely by the top-ranked prediction.

\begin{tcolorbox}[boxrule=0pt,frame hidden,sharp corners,enhanced,borderline north={1pt}{0pt}{black},borderline south={1pt}{0pt}{black},boxsep=2pt,left=2pt,right=2pt,top=2.5pt,bottom=2pt]
    \textbf{Answer to RQ2:} Our analysis reveals statistically significant positive correlations between the \name confidence values and FL performance metrics in both benchmark datasets. Consequently, the confidence value of \name can be used to filter out potentially inaccurate results.
\end{tcolorbox}

\subsection{RQ3: Explanation Quality}

To evaluate the explanations generated by \name, two authors independently rated 300 explanations from Defects4J and then resolved any differences through discussion. To quantify the overall agreement between the authors, we consolidated all labels from multiple attributes into a single list for each author and conducted the agreement analysis accordingly. The agreement of the initial evaluations made was 86.5\%, and the Cohen's $\kappa$ coefficient used by prior work to measure inter-rater agreement~\cite{Wang2021MLUnitTest} was 0.55, a fair to good level of agreement~\cite{fleiss2013statistical}.

The quality evaluation results are presented in \cref{tab:expl_quality}. In 83.7\% of individual runs, \name could successfully generate an explanation; 
in the other cases,
either \name had gone over the LLM token limit or \name had exhausted the function call budget.
Regarding the other measure, among all 300 individual explanations generated by \name, 20\% of them contained correct descriptions about what was causing the bug (Accurate).
Meanwhile, \name would generate at least one inaccurate statement in 26.3\% of all cases (Imprecise). The most common type of explanation was `Bland' (46.7\%), in which \name would generate an explanation that described the test and the covered methods, but did not provide any additional analysis. 
When explanations are aggregated by bug rather than evaluated individually,
at least one accurate explanation was generated for 56.7\% of all bugs, suggesting that with an appropriate selection of explanations, developers could benefit from the explanations of \name more often.

In this regard, there is a correlation between FL confidence explored in RQ2 and the average quality of explanations for a bug, as presented in \cref{tab:expl_quality}, providing hints as to how to selectively present good explanations. There is a modest improvement in all helpful qualities of explanations when filtering them by confidence, while the increase in the harmful quality (Imprecise) is negligible. These results show that there is both promise, and room to improve, when it comes to automatically identifying helpful explanations from the human perspective, a topic we further discuss in \cref{sec:future_dirs}.

\begin{figure}
    \centering
    \includegraphics[width=0.95\linewidth]{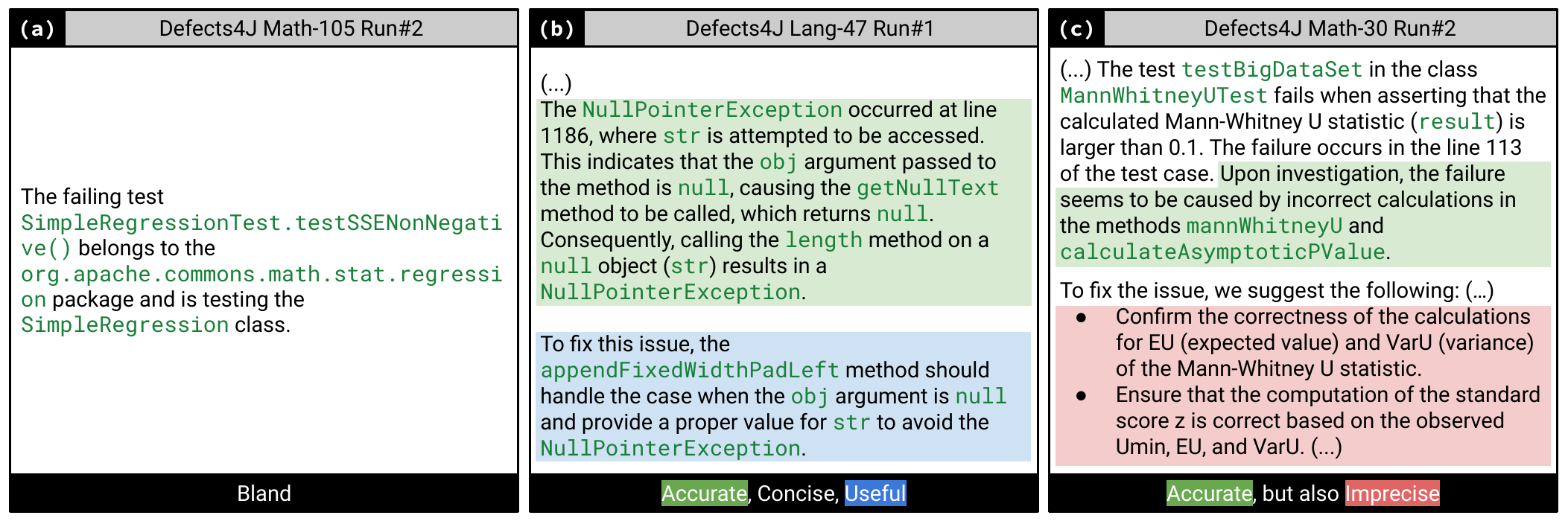}
    \caption{Example explanations from \name. Explanations (b) and (c) are truncated for clarity.}
    \label{fig:category_examples}
\end{figure}

To further clarify these results, we present three \name-generated explanations in \Cref{fig:category_examples}. 
In \Cref{fig:category_examples} (a), the explanation is true, but provides no real information about what the underlying bug is, so it is a `bland' explanation that does not further the interests of the developer. 
Meanwhile, in \Cref{fig:category_examples} (b), we see a good explanation that is simultaneously accurate, concise, and useful. The first part of the explanation (green) accurately describes how the error manifested by detailing which operations culminated in the error. In the next part of the explanation (blue) it goes on to describe what should be done to fix the issue, which corresponds with the actual developer fix. As all of the provided information is likely genuinely helpful and there is no extraneous content, the explanation also qualifies as a concise explanation.
Finally, in \Cref{fig:category_examples} (c), we present an example of an explanation that is accurate but also partially imprecise. For this bug, the buggy method is not immediately called by the failing test, nor visible in the exception call stack. Despite this, this explanation accurately pinpoints the method call chain leading to the bug (green), and thus is accurate as it provides information that could help the developer understand why each method was suggested by \name. However, the fixes suggested by the explanation are imprecise (red), as they deviate from the actual developer fix.
As this imprecise recommendation can be regarded as extraneous content, the explanation is not concise; nor is it useful, as the suggested fix is wrong.

To understand the circumstances in which \name failed to generate good explanations, we analyzed the 26 bugs for which no accurate explanations were generated. Overall, we found that there were four main causes of failure. In 14 cases (53.8\% of failing bugs), we found that their tests relied on custom test helper functions, particularly for the bugs from the Closure project, which has non-conventional test cases~\cite{martinez2017automatic}. As the LLM of \name was generally unaware of the precise semantics of these test cases, it spent most of its time retrieving the helper functions, and less time inspecting potentially buggy code. This suggests when using \name, ideally tests should be self-contained; further research is required to incorporate project-specific information effectively. In the second scenario, \name failed for six bugs as the tests used too many classes and methods for \name to effectively inspect within the $N=10$ function call budget. In such cases, techniques such as test case purification may also help reduce the search space~\cite{Xuan2014TestPurification}. As for the remainder, in three cases, the buggy methods were so long that they eventually caused a length error for \name; meanwhile, in three cases the LLM consistently made logical mistakes. The steady improvement of LLMs may help deal with these problems - indeed, GPT-4 could make correct explanations for three of these six bugs (one for the context length limit and two for the logical mistakes).

\begin{table}[t]
\caption{Explanation rating results of \name-GPT3.5}
\centering
\scalebox{0.8}{
\begin{tabular}{lrrrrrrr}
\toprule

Subset & Exists & Accurate & Imprecise & Concise & Useful & `Bland' & Total\\\midrule
\textbf{Individual Explanations} & 83.7\% & 20.0\% & 26.3\% & 9.3\% & 8.0\% & 43.0\% & 300\\
$0.00 \leq $ Confidence $ < 0.25 $ & 78.3\% & 10.0\% & 24.2\% & 3.3\% & 1.7\% & 46.7\% & 120\\
$0.25 \leq $ Confidence $ < 0.50 $ & 87.5\% & 23.8\% & 28.8\% & 7.5\% & 11.3\% & 43.8\% &  80\\
$0.50 \leq $ Confidence $ < 0.75 $ & 81.5\% & 26.2\% & 24.6\% & 16.9\% & 12.3\% & 36.9\% &  65\\
$0.75 \leq $ Confidence $ \leq 1.00 $ & 97.1\% & 34.3\% & 31.4\% & 20.0\% & 14.3\% & 40.0\% & 35\\ \midrule
\textbf{Aggregated By Bug} & 100\% & 56.7\% & 66.7\% & 31.7\% & 23.3\% & 93.3\%  & 60\\
$0.00 \leq $ Confidence $ < 0.25 $ & 100\% & 37.5\% & 70.8\% & 16.7\% & 8.3\% & 95.8\% & 24\\
$0.25 \leq $ Confidence $ < 0.50 $ & 100\% & 62.5\% & 68.8\% & 31.3\% & 31.3\% & 93.8\% &  16\\
$0.50 \leq $ Confidence $ < 0.75 $ & 100\% & 69.2\% & 53.8\% & 46.2\% & 30.8\% & 84.6\% &  13\\
$0.75 \leq $ Confidence $ \leq 1.00 $ & 100\% & 85.7\% & 71.4\% & 57.1\% & 42.9\% & 100\% & 7\\
\bottomrule
\end{tabular}}
\label{tab:expl_quality}
\end{table}

\begin{tcolorbox}[boxrule=0pt,frame hidden,sharp corners,enhanced,borderline north={1pt}{0pt}{black},borderline south={1pt}{0pt}{black},boxsep=2pt,left=2pt,right=2pt,top=2.5pt,bottom=2pt]
    \textbf{Answer to RQ3:} About 20\% of \name explanations accurately describe the root cause of the bug, per a manual assessment; for 56.7\% of all bugs, an accurate explanation is generated at least once. Interestingly, helpful explanations are more common for bugs when \name is confident.
\end{tcolorbox}

\subsection{RQ4: Developer Feedback}

Finally, we present a summary of the feedback we received from our semi-structured interviews of developers, who debugged real-world bugs from the \texttt{pandas} project, sourced from BugsInPy. In this section, we showcase the common answers from developers on seven key questions; 
further responses and analysis can be found in the supplementary material~\cite{autoflsupplementary}. In this section, we refer to individual developers using their anonymous animal IDs used in our study, e.g., \texttt{seal}.

\emph{\textbf{Is FL Wanted?}} We asked if developers wanted to use FL; in our study, we had to clarify that this meant being provided a list of suspicious code elements without explanations. Of the 16 developers, 13 agreed or conditionally agreed that FL (even without explanations) would help their debugging efforts. Several developers remarked that the utility of FL would depend on their familiarity with a project, noting that FL would be particularly helpful when the developer is unfamiliar with the subject system, and less so when the developer has intimate knowledge. A small number of developers disagreed that FL would help, as they were confident that they could perform FL based on the error message. This generally positive attitude, albeit with conditions and reservations, is similar to what was reported by Kochhar et al.~\cite{Kochhar2016FLExpect} in their survey of developer expectations on FL.

\emph{\textbf{Are FL Explanations Wanted?}} Critically for our work, we asked developers whether they wanted explanations for FL. Four developers described explanations for FL as necessary, while eight additionally described explanations as useful; in total, twelve developers suggested that they would want explanations when using FL. Developers commonly believed that explanations would help them navigate unfamiliar code, and help them think through or fix the bug; for example, \texttt{parrot} noted that ``with just a location suggested, it's vague how to fix the bug; with the cause of the bug explained, it was easier to think''. Meanwhile, developers also expressed concern about explanations, which were generally conditional: one developer worried about the accuracy of explanations on difficult bugs, and four thought that explanations would be unnecessary for easy code. On the flip side, this shows that every developer agreed that explanations would be helpful in unfamiliar projects or difficult bugs, as long as the explanation is reasonably accurate.

\emph{\textbf{What were the strengths of explanations from \name?}} Based on the explanations suggested in our experiments, we asked developers to describe the strengths of the explanations generated by \name. Developers appreciated that the explanations described the intention of the function(s) under test, and (when fixes were provided) generally liked the fixes suggested by \name. For example, \texttt{koala} noted that ``the error message alone didn't provide much of a starting point; by following and explaining the execution context of the bug, it was more convenient to solve the debugging problems''. The most common theme was that developers liked a natural language explanation of the error message, with five developers liking this aspect - for example, \texttt{turtle} described the explanations as ``less clunky when compared to terminal messages, and more human-friendly; even if it ultimately takes longer to read the message, it felt better.''

\emph{\textbf{What was unnecessary in the explanations of \name?}} However, the same feature was also seen as unnecessary as well. Two developers thought that the error message explanation did not aid their understanding of the bug, suggesting customization or compartmentalization in explanations could help improve user experience in general; we discuss this issue in greater depth in our discussion of the ideal explanation. Among developers who said that there was unnecessary content, most said that the overlapping content in the explanations (both between the ten explanations that we provided per bug, and within each explanation) was unnecessary, and that it would be helpful to summarize explanations with the same content. One suggestion was that clustering similar explanations together would be another way of improving developer consumption of explanations. Outside of the overlapping content, most developers (10) did not find anything to remove from the explanations; they often noted that unnecessary content could be quickly skipped as well, indicating that it was not a significant problem.

\emph{\textbf{What was confusing in the explanations of \name?}} As the explanations are generally presented in a confident tone, there is a risk that developers could be confused by inaccurate explanations. Indeed, while six developers found nothing confusing about the explanations, four others thought that the suggestions of the explanations to fix tests were confusing, and two developers noted that the patches or fault locations suggested by the explanations were insufficient to fix the bug, leading to confusion and wasting their time.

\emph{\textbf{What is the ideal explanation?}} We asked developers about what their ideal explanation of a bug or FL results would be, to identify how the explanations of \name could improve. Five developers noted that having a clear template for explaining the bug and the fix would have helped; for example, \texttt{koala} suggested a tripartite template which would clearly show ``where the bug happened, why this is likely problematic behavior, and how to fix the bug''. Such a template would also help accommodate the most commonly identified ideal features in a bug explanation - (i) the logic of the failure, requested by five developers; (ii) the original intention of the code/test, requested by four developers; (iii) and finally a suggested patch, which was included in the ideal explanation of 13 developers and was by far the most popular feature in an ideal bug/FL explanation. Furthermore, four developers suggested that having real dynamic values incorporated in the explanation would help them understand the flow of the bug: \texttt{koala} noted that ``I had to keep track of what the variable values were for the bug-revealing test manually; if this information were integrated into the explanation, that would be more convenient''. Finally, when asked about the ideal number of explanations presented to the developer, 11 developers thought that the ten explanations presented in the experiment were too much, and wanted a summary. Nonetheless, they were open to seeing multiple explanations - \texttt{chicken} noted that ``as long as the content is different, having multiple options to choose from is good as well''. In this context, \texttt{dragon} suggested that ideally, explanations would have a confidence value, which would help developers choose between multiple explanations.%

\emph{\textbf{In what order did developers read the explanations?}} 
We asked developers whether they had any heuristics on how to choose which explanation to read. While many answers were given, we focused on the two most common answers. First, four developers answered that they had looked at the FL list, and looked at explanations that supported the most likely location first. 
On the other hand, seven developers answered that they simply looked at the explanations from top to bottom, without consulting the FL list or any other information. This suggests that it would likely improve user experience and debugging time if the quality of explanations could be automatically assessed, and high-quality explanations could be presented first.

\begin{tcolorbox}[boxrule=0pt,frame hidden,sharp corners,enhanced,borderline north={1pt}{0pt}{black},borderline south={1pt}{0pt}{black},boxsep=2pt,left=2pt,right=2pt,top=2.5pt,bottom=2pt]
    \textbf{Answer to RQ4:} Developers were generally supportive of explainable FL, and suggested that the explanations of \name were helpful by explaining the error message and the intention of each function. Meanwhile, developers disliked the overlap in content in the presented explanations, and found some explanations inaccurate. Ideally, developers wanted templated explanations that would help them quickly find what they wanted, such as the bug logic or patch.
\end{tcolorbox}

\section{Future Directions: Better Quality Estimation for \name}
\label{sec:future_dirs}

The results of RQ4 make clear the need for selecting reliable bug explanations. The natural follow-up question is whether there are any features that can predict the quality of explanations. Explanations, being long-form text, are more difficult to automatically evaluate: they are not directly executable, nor is it possible to use self-consistency~\cite{wang2023selfconsistency} to check which are good explanations, as explanations can syntactically differ while having similar semantic content. One way of tackling this is to get downstream artifacts that can be subjected to evaluation using self-consistency, such as FL predictions; as demonstrated in RQ2, there is a correlation between FL confidence and explanation quality. On the other end, we may make executable downstream artifacts, and see whether explanations with high-quality executable artifacts are generally of higher quality themselves.
In this section, we explore this question by presenting the preliminary results of such a prediction task on the explanations of \name, based on the following features:

\begin{itemize}[leftmargin=15pt]
    \item \textbf{Test Score}: Based on a bug explanation from \name, GPT-3.5 is repeatedly prompted to generate (i) a failing test that reproduces the bug and (ii) a passing test similar to the bug-reproducing test. The explanation is scored by the ratio of tests that behave as expected (i.e., the generated bug-reproducing test should fail). The intuition is that a good explanation of the bug could allow an LLM to generate bug-reproducing tests~\cite{Kang2023aa}.
    \item \textbf{APR Score}: Based on a bug explanation and the code of buggy methods suggested by \name, GPT-3.5 is repeatedly prompted to generate a patch that would fix the bug. The explanation is scored by the ratio of partial fixes (e.g., the ratio of patches that make at least one previously failing test pass). The intuition is that a good explanation of the bug could ease patch generation when using an LLM.
    \item \textbf{GPT Scores}: GPT-3.5 is prompted to rate an explanation from \name on the same four quality measures (Accurate, Imprecise, Concise, Useful) that we used in RQ3 on a five-point Likert scale. The intuition is that an LLM may be capable of `reflection' on its own answers~\cite{shinn2023reflexion}, and be capable of correctly identifying which explanations are accurate.
    \item \textbf{Explanation Length}: An explanation is scored by its length. We include this as a simple baseline to evaluate the validity of any results from the aforementioned metrics.
\end{itemize}

The non-parametric Spearman's rank correlation coefficient between these features and the quality of explanations as evaluated in RQ3 is presented in \cref{tab:expl_corr}. First, among the dynamic features, the test score shows a statistically significant correlation with positive explanation qualities such as `Accurate', while showing a low correlation for `Wrong'. APR score shows similar correlation characteristics, but on a smaller magnitude, as partial patches were rarer (55 explanations from our manual examination led to a partial patch), and a slightly negative correlation with wrong explanations. Overall, while dynamic features show promise in identifying truthful explanations, their correlation as of now is not strong enough to fully rely on. 
More detailed tables that present bug-aggregated and bug-controlled correlation values are presented in the supplementary material~\cite{autoflsupplementary}.

Among the GPT Scores, only $\text{GPT}_\text{useful}$ showed interesting trends; it showed a significant correlation ($p<0.001$) with every quality, including `Wrong', suggesting that GPT-3.5 will rate any explanation that is detailed as useful. The Length feature provides a useful comparison as well: like $\text{GPT}_\text{useful}$, Length shows a significant correlation with every quality (except `Useful'), and the correlation between Length and $\text{GPT}_\text{useful}$ was also significant at 0.49. As a result, further research needs to be conducted to make pure LLM-based evaluation of LLM-generated explanations viable.

\begin{table}[t]
\caption{Spearman Correlation between explanation quality predictors and actual quality. Results with $p<0.01$ are marked with *, and results significant with $p<0.001$ are marked with **.}
\scalebox{0.8}{
\begin{tabular}{lllll}
\toprule
            name &  Test Score & APR Score & $\text{GPT}_\text{useful}$ & Length \\
\midrule
        Accurate        &      $+0.2358$** &   $+0.1946$*  &    $+0.3759$** &  $+0.3009$** \\
`Wrong' (only imprecise)&      $+0.0408$   &   $-0.0643$   &    $+0.3266$** &  $+0.3271$** \\
          Useful        &      $+0.2635$** &   $+0.1942$*  &    $+0.2371$** &  $+0.1585$   \\
         `Bland'        &      $-0.2364$** &   $-0.1105$   &    $-0.6026$** &  $-0.5391$** \\
     FL Accurate        &      $+0.2737$** &   $+0.4923$** &    $+0.1437$   &  $+0.1528$   \\
\bottomrule
\end{tabular}
}
\label{tab:expl_corr}
\end{table}

\begin{figure}[t]
    \centering
    \begin{subfigure}[t]{0.45\linewidth}
        \centering
        \includegraphics[width=\linewidth]{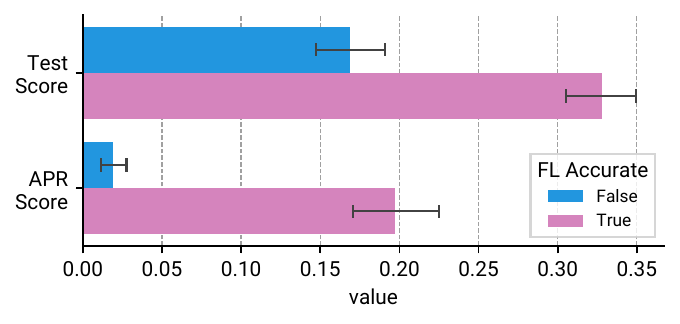}
        \caption{Mean of test and APR scores}
        \label{fig:test_apr_score}
    \end{subfigure}
    \begin{subfigure}[t]{0.45\linewidth}
        \centering
        \includegraphics[width=\linewidth]{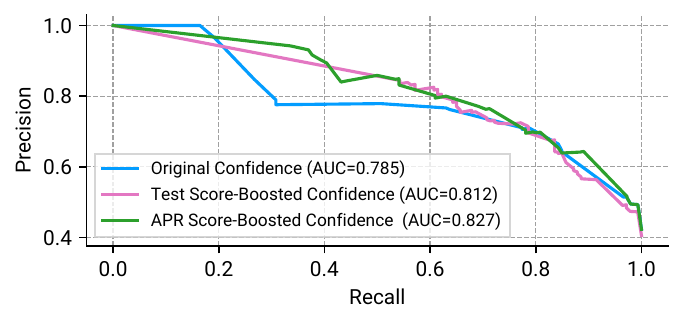}
        \caption{Precision-recall curve for P@1 prediction}
        \label{fig:conf_pr_boosted}
    \end{subfigure}
    \caption{Relationship between dynamic scores and FL performance.}
    \label{fig:boost}
\end{figure}

During the evaluation, the significant correlation between both of the dynamic scores (Test Score and APR score) and FL accuracy (bottom row of \cref{tab:expl_corr}, \cref{fig:test_apr_score}) piqued our interest: could dynamic scores be used to more accurately \textit{predict} \name's performance? To explore this, we conducted experiments that boosted scores of methods using the test score and APR as in \cref{eq:boost_score}, where $boost_{i}$ is either the \textbf{test score} or the \textbf{APR score} of the $i$-th explanation. Intuitively, if an \name run has a higher dynamic score, the methods predicted in that run get a greater boost. Using APR to boost FL bears similarities to previous debugging work, such as MUSE~\cite{MUSE2014Moon} or ProFL~\cite{lou2020can}, while the combination of test generation and FL has not yet been explored to the best of our knowledge.
\begin{equation}
score_{new}(m) = min(1, score(m) \times \prod_{\{1 \leq i \leq R|m \in r_i\}}\left(1 + boost_{i}\right))
\label{eq:boost_score}
\end{equation}

\cref{fig:conf_pr_boosted} presents the precision-recall curve for predicting Precision@1 based on the confidence thresholding on the Defects4J dataset. 
The new confidence values which incorporate dynamic scores are more accurate estimators of Precision@1 than the original confidence score, with the area under the curve (AUC) increased by up to 5.4\% using the APR score. This shows that dynamic features, i.e., the performance of the downstream tasks, can help refine \name results as well.

\begin{tcolorbox}[boxrule=0pt,frame hidden,sharp corners,enhanced,borderline north={1pt}{0pt}{black},borderline south={1pt}{0pt}{black},boxsep=2pt,left=2pt,right=2pt,top=2.5pt,bottom=2pt]
    \textbf{Future Directions:} Preliminary results show that dynamic evaluation of explanations has the potential to identify both helpful explanations and accurate FL.
\end{tcolorbox}

\section{Threats to Validity}
\label{sec:ttv}

\noindent{\textbf{Internal Validity}} The computation time of \name varies, as it relies on OpenAI server conditions and the random nature of LLMs. To mitigate this, the time cost of \name was averaged over multiple runs. In evaluating the quality of explanations generated by \name, human error in the validation process is possible. To address this, two authors independently assessed explanations and resolved disagreements. Data leakage is a concern with LLMs, e.g., the possibility of bug-fixing commits being contained in their training data. However, the comparison results of \name with the Test-GPT3.5 baseline in RQ1, which shares the same model but does not interact with the functions, suggest that \name's performance is not solely due to model memorization. 

\noindent{\textbf{Construct Validity}} To gather developer feedback, we engaged with professional developers to assess their experience with \name. Due to security reasons, these developers tested \name on an open-source project (\texttt{pandas}) rather than the projects that they are working on. Consequently, their responses and impressions may not entirely reflect the experiences that developers would have during real-world debugging in their work projects.

\noindent{\textbf{External Validity}} 
Our evaluation of \name primarily focused on programs with unit tests from Java and Python. While our findings can be generalized within these contexts, they may not extend to other programming languages or different levels of testing.
While we interviewed developers from three IT companies to gather insights into their experiences with \name, their responses may not be fully representative of all developers, given the diverse perspectives in the software industry. Furthermore, the performance of \name can be affected by various factors, such as the choice of the language model. 

\section{Conclusion}
\label{sec:conclusion}

This paper presents \name, an explainable LLM-based FL technique that has many useful characteristics that make it easier for practitioners to adopt, with particular strengths being its low requirement of software artifacts (only a single failing test is required), reasonable runtime, and critically its ability to generate explanations. Our evaluation shows the strong performance of \name as a standalone FL tool, with \name outperforming the baselines that we compare against. The explanations generated by \name required more nuanced evaluation: manual evaluation of the explanations revealed \name could generate an accurate explanation for 56.7\% of all bugs. Surveying developers on what they wanted in bug explanations, 
many developers called for structured explanations that explain the intention of tests, how the bug happened, and how to fix a bug; furthermore, they were willing to see only a few explanations. Motivated by this need to pick explanations, we present preliminary results on automatically identifying high-quality explanations suggesting that while dynamic features show promise, further research is necessary. Based on these results, we hope to continue investigating how to consistently make useful explanations of bugs, which is one of the unique ways in which LLMs could benefit practitioners. %

\section*{Data Availability}
Our source code and data are publicly available at both GitHub~\cite{autofl_github} and Figshare~\cite{autofl_artifact} (archived).

\section*{Acknowledgements}
We would also like to thank the anonymous reviewers for their thorough and helpful comments. This work was supported by the National Research Foundation of Korea (NRF) funded by the Korean Government MSIT (RS-2023-00208998), the Engineering Research Center Program (2021R1A5A1021944), as well as the Institute of Information \& Communications Technology Planning \& Evaluation (IITP) grant funded by the Korea government (MSIT) (2021-0-01001).

\bibliographystyle{ACM-Reference-Format}
\bibliography{sample-base}

\end{document}